# Evidence of spin density wave in LaFeAsO, the parent material of the new Fe-based oxypnictide superconductors

**Motoyuki Ishikado, Ryoichi Kajimoto, Shin-ichi Shamoto,**

**Masatoshi Arai***

**Japan Atomic Energy Agency, 2-4 Shirakata-Shirane, Tokai, Ibaraki, 319-1195, Japan**

**Akira Iyo, Kiichi Miyazawa, Parasharam M. Shirage, Hijiri Kito,**

**Hiroshi Eisaki**

**Nanoelectronic Research Institute, National Institute of Advanced Industrial Science and Technology (AIST), Tsukuba Ibaraki 305-8568, Japan**

**SungWng Kim, Hideo Hosono**
**Tokyo Institute of Technology, 4259 Nagatsuta, Midori-ku, Yokohama 226-8503, Japan**

**Tatiana Guidi, Robert Bewley, Stephen M. Bennington**

**ISIS facility, Ruther Appleton Laboratory, Didcot, Oxfordshire, UK**

The Fe-based Oxypnicide[1] superconductors have generated a huge amount of interest; they are a high temperature superconductor, with a $T_c$ of 55K [2], but do not have the two dimensional copper oxygen layer that was thought essential for superconductivity at these high temperatures. Initial studies have hinted towards the possibility of a spin density wave (SDW) in these compounds and how they could play an important role in the superconductivity. SDW's occur at low **temperatures in low-dimensional materials with strong electron correlations or in metals with a high density of states at the Fermi surface. A transition to the SDW state has many similarities to the superconducting transition, driven by the condensation energy with an energy gap opening, and in many materials the SDW state occurs adjacent in the phase diagram to the superconducting state. In these materials it is believed that electron doping suppresses the SDW instability allowing superconductivity to emerge. In this paper we report on the first direct experimental evidence of a SDW in the LaFeAsO parent compound as observed with inelastic neutron scattering. We show that these excitation derive from a two-dimensional Fermi surface nesting with a nesting vector of ($\pi,\pi,0$)**

The discovery of the Fe-based oxypnictide superconductors [1] have generated a huge amount of interest and a race to pin down the mechanisms of the superconductivity through various experimental and theoretical techniques. Despite more than 20 years of work the mechanisms that drive high-temperature superconductivity are still controversial. These new materials are based on a two dimensional FeAs layer rather than the $CuO_2$ layer found in the perovskite superconductors, so for the first time this opens up the opportunity to compare two classes of high-temperature superconductor based on different parent compounds.

Thermodynamic and transport measurements show that the parent compound, the Fe-oxypnictide LaFeAsO, has a structural phase transition at $T_s$ = 155K [3], which has been associated with the formation of a SDW state with the Fermi-surface nesting vector ($\pi, \pi, 0$) [4]. The structural transition is from tetragonal to monoclinic; and at temperatures below $T_S$, Cruz et al. have observed the growth of peaks at $(1,0,1)_M$, where the magnetic cell is defined by $\sqrt{2}a \times \sqrt{2}b \times 2c$ as defined in Ref.[3], and $(1,0,3)_M$ confirming the occurrence of anti-ferromagnetic long range order with a Neel temperature of $T_N$~135K. Although it is possible that this ordering is related to a SDW such evidence is not definitive. In this letter we report on inelastic neutron scattering measurements which show conclusively that a SDW exists in the parent compound.

We performed inelastic neutron scattering measurements, on the MERLIN spectrometer in the ISIS facility in the Rutherford Appleton Laboratory, UK [5]. The measurements were on a 34 gram powder sample of the parent compound LaFeAsO at temperatures ranging from 7K to room temperature. Figure 1 shows the two-dimensional intensity plots of the dynamical structure factor, S(Q,E), at 7K, 140K and 240K observed with an incident energy of 30meV. At 140K, very close to $T_N$, we observe a prominent vertical line at rising from a momentum transfer of 1.1 Å$^{-1}$; close to the (1,0,1)$_M$ reflection point reported in neutron diffraction results of Cruz et al.[3]. In order to estimate the magnetic signal, phonon intensity estimated from the 240K data with correction on the Bose population factor was subtracted from the 140K data and shown in Fig.1 (d). A vertical excitation at 1.1 Å$^{-1}$ can clearly be seen in the figure, and another line is even seen rising from 2.5 Å$^{-1}$. The excitations seem to extend as high as 20meV.

In order to clarify the nature of the excitation, slices of S(Q,E) constant energy for various temperatures, integrated between 7.5meV and 9meV where there are no obvious optic phonons, are shown in figure 2. At T$_N$ prominent peaks occur at 1.1 Å$^{-1}$ and 2.5 Å$^{-1}$. At temperatures above and below T$_N$ these peaks loose intensity and become broader. At 7K the magnetic scattering seems to be very weak. In order to estimate the magnetic signal we scaled the 7K data by the Bose population factor and subtracted it from data (figure 3). The resultant peak shape is not symmetric and has a tail at the higher Q side, which is what one would expect from the scattering of a powder sample with a two-dimensional atomic structure, i.e. a Bragg rod in the reciprocal space [6]. Figure 4 shows the integrated intensity of the magnetic scattering as a function of temperature. It is clear that the intensity has a maximum around T$_N$ and a temperature evolution that closely mirrors the behaviour of the resistivity. The weak temperature dependence above T$_N$ suggests that the system is 2-dimensional with strong correlation in the layer[7].

The present superconductor has 2-dimentional layers composed from interpenetrating iron and arsenic square lattices. The calculated electronic band structure shows a 2-dimensional Fermi surface nesting between a hole-like Fermi surface at the $\Gamma$-point and an electron-like Fermi surface at the M-point, with a characteristic nesting vector of ($\pi$, $\pi$, 0) [4, 8]. The present magnetic excitation with a maximum intensity at T$_N$ occurs very close to the antiferromagnetic Bragg peak positions at Q=(1,0,1)$_M$ =1.16 Å$^{-1}$ and (1,2,1)$_M$=2.49 Å$^{-1}$ [3]. These positions are very close to the expected nesting vector at Q=($\pi$, $\pi$, 0)=1.10 Å$^{-1}$, and (3$\pi$, $\pi$, 0)=2.46 Å$^{-1}$ and as such cannot be distinguished between each other because of insufficient momentum transfer resolution in the present measurement. Here, we emphasize that we could not observe any excitation at around (1,0,3)$_M$ =1.54 Å$^{-1}$ within the experimental accuracy, despite the fact that magnetic Bragg intensity was observed indicating antiferromagnetic long range order. This is consistent with excitations from a SDW caused by nesting of the 2-dimentional Fermi surface that only appears at the reciprocal points of odd multiples of ($\pi$, $\pi$, 0), such as (3$\pi$, $\pi$, 0), ($\pi$, 3$\pi$, 0) and so on [4, 8]. Therefore, absence of the excitation at (1,0,3)$_M$ is conclusive evidence of the excitation from a SDW and not a spin wave excitation from long range antiferromagnetic correlations, which should exist at all antiferromagnetic Bragg points. The peak intensities around Q=($\pi$, $\pi$, 0) and (3$\pi$, $\pi$, 0) have comparable intensities and this is reasonable if we take the magnetic form factor of Fe$^{2+}$ and multiplicity of those peaks into account.

The temperature evolution of the excitation intensity with a maximum at around T$_N$ (~140K) is similar to the canonical three-dimensional SDW observed in chromium, and can be characteristics of SDW [9]. The persistence of the intensity above T$_N$ also suggests the interactions are 2-dimensional in the FeAs layer in contrast to the three-dimensional interaction in chromium. An abrupt drop in intensity below T$_N$ is reminiscent of the rapid decrease of the resistivity seen in LaFeAsO [4], suggesting an existence of a strong spin scattering to conducting carriers.

These results provide the first strong experimental evidence that the excitations observed derive from a two-dimensional Fermi surface nesting with the nesting vector at ($\pi$, $\pi$, 0) as was suggested in previous reports [4, 8].

**Acknowledgement**

We acknowledge the ISIS facility for accepting our urgent proposal at short notice and the indispensable support for the experiment. The experiment was done under UK-Japan collaboration and a support of the grant-of–aid in the specially promoted area, 17001001.

**References**

[1] Kamihara, Y., Watanabe, T., Hirano, M., & Hosono, H. Iron-based layered superconductor La[O$_{1-x}$F$_x$]FeAs ($x$ = 0.05-0.12) with $T_c$ = 26 K. *J. Am. Chem. Soc.* **130,** 3296 (2008).

[2] Zhi-An, R. *et al.* Superconductivity at 55 K in iron-based F-doped layered quaternary compound Sm[O$_{1-x}$F$_x$] FeAs. *Chinese Phys. Lett.* **25**, 2215 (2008).

[3] De la Crutz, C. *et al.* Magnetic order close to superconductivity in the iron-based layered LaO$_{1-x}$F$_x$FeAs systems. *Nature* **453**, 899 (2008).

[4] Dong, J. *et al*. Competing orders and spin-density-wave instability in La(O$_{1-x}$F$_x$)FeAs. *Europhys. Lett.* **83**, 27006 (2008)

[5] Bewley, R. I . *et al*. MERLIN, a new high count rate spectrometer at ISIS. *Physica B* **385-386**, 1029 (2006).

[6] Warren, B.E. *Phys. Rev.* **59,** 693 (1941).

[7] Fisher, M. E. Lattice statistics in a magnetic field. I. a two-dimensional super-exchange antiferromagnet. *Proc. Roy. Soc.* **A254**, 66 (1960).

[8] Singh, D. J. & Du, M.-H., Density functional study of LaFeAsO$_{1-x}$F$_x$: a low carrier density superconductor near itinerant magnetism. *Phys. Rev. Lett.* **100**, 237003 (2008)

[9] Fawcett, E. Spin-density-wave antiferromagnetism in chromium. *Rev. Mod. Phys.* **60**, 209 (1988).

Fig.1 The dynamical structure factor measured at 240K (a), 140K(b), 7K(c) and estimated magnetic signal at 140K by subtracting the phonon contribution scaled from 240K data (d). The incident energy is Ei=30meV. A steeply dispersing excitation can be seen rising from the position at 1.1 Å$^{-1}$ and 2.5 Å$^{-1}$ which persist up to at least 20meV.

Fig.2: Constant energy cuts taken from S(Q,E) measured at 30meV and integrated between 7.5meV and 9meV. This is well below any of the obvious optical phonon modes. Peaks can be seen to develop around $T_N$=140K close to the ($\pi$, $\pi$, 0) and ($3\pi$, $\pi$, 0) Fermi surface nesting vectors at 1.1 Å$^{-1}$ and 2.5 Å$^{-1}$.

Fig.3: Estimated magnetic signal at various temperatures by subtracting the data at 7K scaled for the Bose population factor. Each of data is shifted by 0.4 in intensity. The shape of the excitation at 1.1 Å$^{-1}$ has a long tail at higher momentum transfers, as would be expected for two-dimensional excitation. The line is the fit curve using a function in Ref. [6]

Fig.4 The temperature dependence of the SDW intensity integrated between 7.5meV and 9meV and between 1.0 and 1.3 Å$^{-1}$. A maximum can be seen at $T_N$=140K, as expected for a SDW excitation. An abrupt decrease below $T_N$ resembles the critical behaviour of the resistivity.

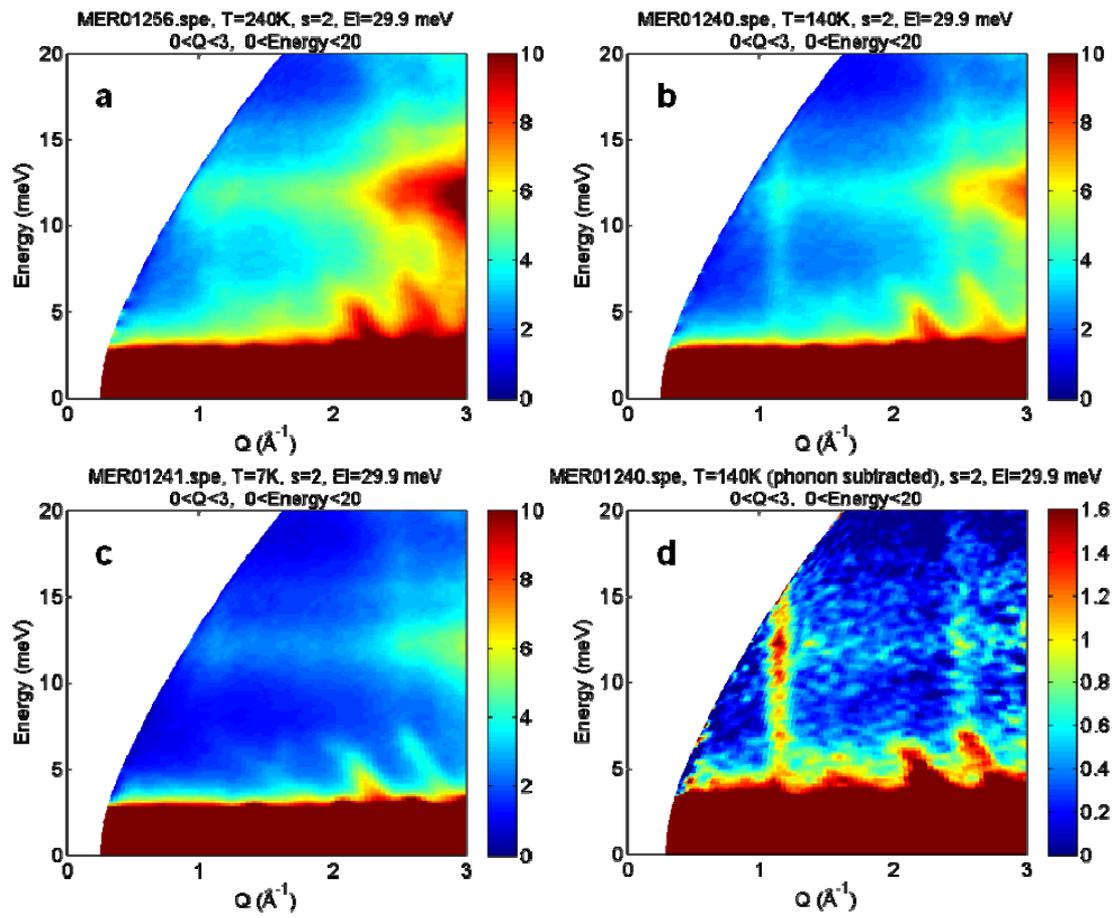

Fig. 1

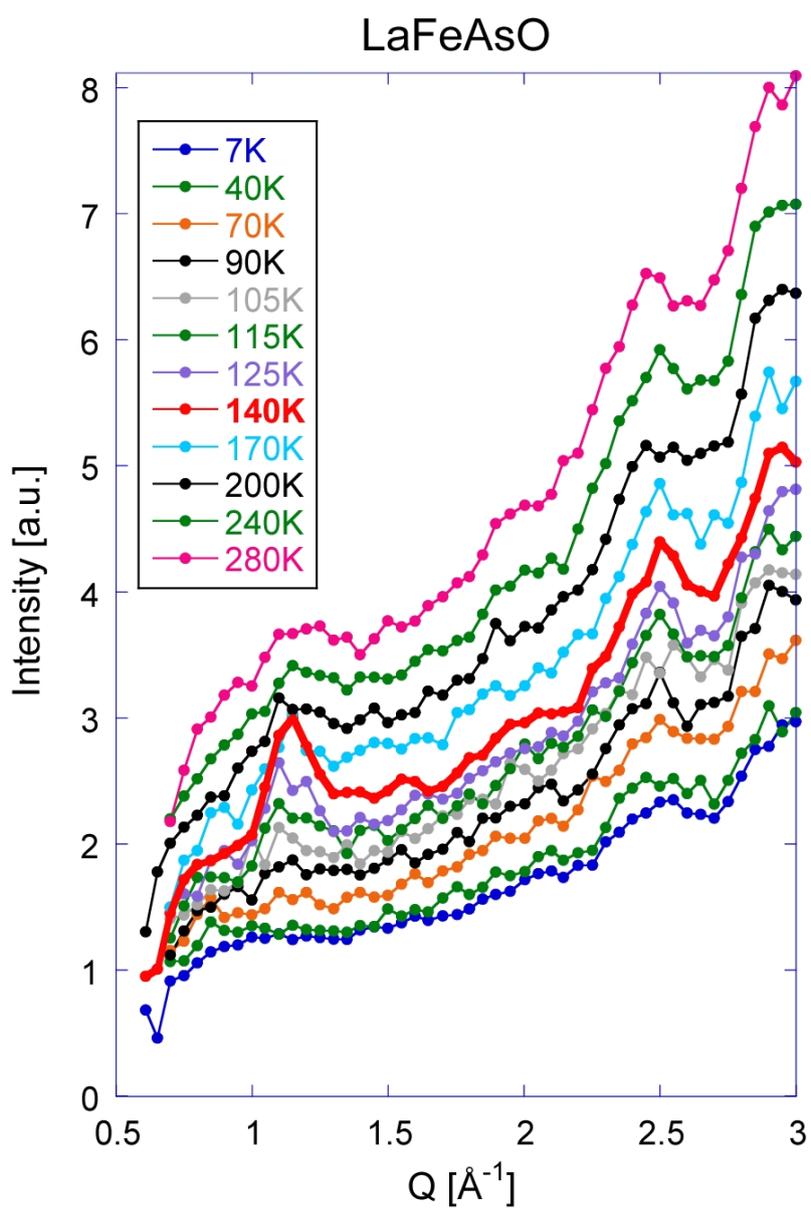

Fig. 2

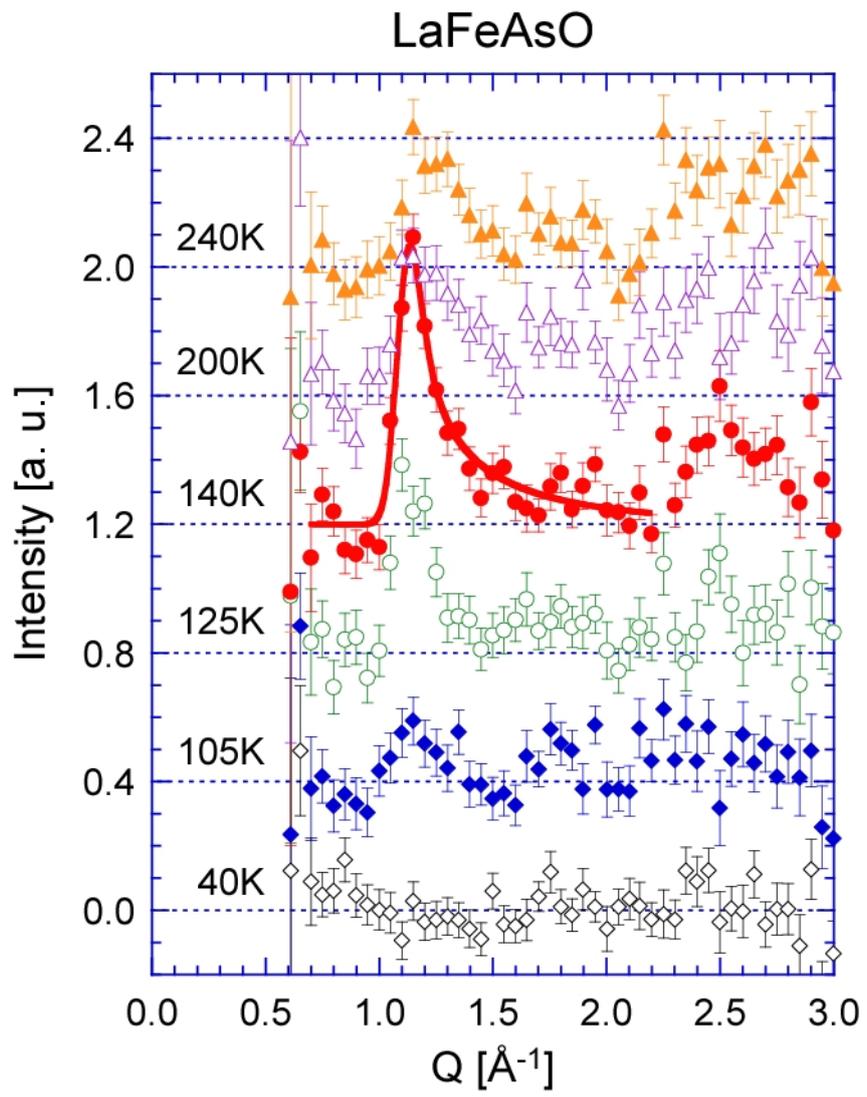

Fig. 3

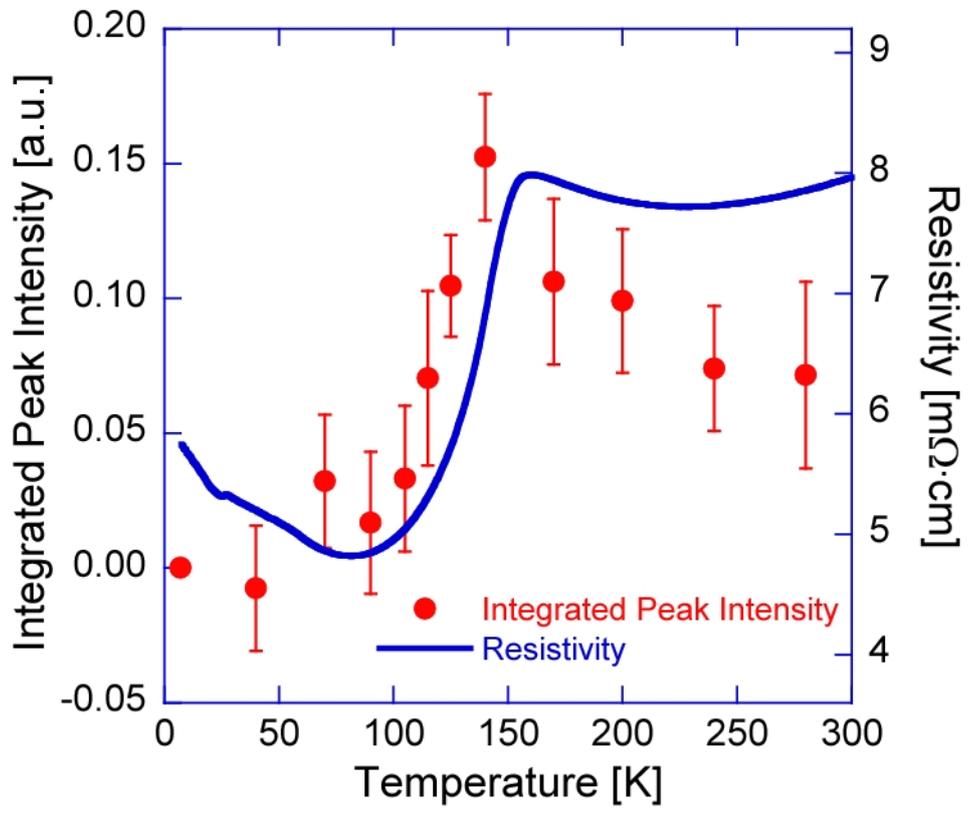

Fig. 4